# Test of Cosmic Spatial Isotropy for Polarized Electrons Using a Rotatable Torsion Balance

Li-Shing Hou , Wei-Tou Ni and Yu-Chu M. Li
Center for Gravitation and Cosmology
Department of Physics, National Tsing Hua University
Hsinchu, Taiwan 30055, Republic of China

**Abstract**

To test the cosmic spatial isotropy, we use a rotatable torsion balance carrying a transversely spin-polarized ferrimagnetic $Dy_6Fe_{23}$ mass. With a rotation period of one hour, the period of anisotropy signal is reduced from one sidereal day by about 24 times, and hence the $1/f$ noise is greatly reduced. Our present experimental results constrain the cosmic anisotropy Hamiltanian $H = C_1\sigma_1 + C_2\sigma_2 + C_3\sigma_3$ ($\sigma_3$ is in the axis of earth rotation) to $(C_1^2 + C_2^2)^{1/2} = (1.8 \pm 5.3) \times 10^{-21}$ eV and $\mid C_3 \mid = (1.2 \pm 3.5) \times 10^{-19}$ eV. This improves the previous limits on $(C_1, C_2)$ by 120 times and $C_3$ by a factor of 800.

PCAS number(s) : 04.80.-y, 11.30.Cp, 98.80.-k

Einstein equivalence principle (EEP) is the cornerstone of metric theories of gravity and governs the microscopic and macroscopic structures in external gravitational fields. In metric theories (including general relativity), the EEP guarantees local Lorentz invariance (LLI). However, in the spirit of Mach, the inertial and related properties should be determined by the distribution of matter in the cosmos. To test this, Hughes-Drever type experiments have been performed over last 40 years with increasing precision on the anomalous atomic energy level splittings of Li [1-3], Be [4-6] and Hg [7-9]. With the advent of the concept of spontaneous broken symmetry of vacuum and the discovery of the quadrupole anisotropy in cosmic microwave background radiation [10, 11], the test of cosmic isotropy at the fundamental law level and the enhancement of precision of the Hughes-Drever type experiments become even more significant.

Ref. [1-9] are mainly Hughes-Drever type experiments on nuclei. Phillips worked on a Hughes-Drever type experiment on electron since 1965 [12]. He used a cryogenic torsion pendulum carrying a transversely polarized magnet with superconducting shields. In 1987 [13], he set a stringent upper limit of $8.5 \times 10^{-18}$ eV for the energy splitting of electron spin-states. In our laboratory we have used a room-temperature torsion balance with a magnetically-compensated $DyFe_3$ polarized-mass to improve the limit. Our cumulated results set a limit of $2.96 \times 10^{-18}$ eV [14-16]. Berglund et al. [9] have used the relative frequency of Hg and Cs magnetometers to monitor the potential energy level variations due to spatial anisotropy and gives an upper limit of $1.7 \times 10^{-18}$ eV for electron. For all of these experiments, the signals detected have period of one sidereal day (23 hr 56 min 4 sec). Table 1 lists the corresponding limits given by various experiments.

For the analysis of cosmic anisotropy for electrons, we use the following Hamiltonian:

$$H = C_1\sigma_1 + C_2\sigma_2 + C_3\sigma_3 \quad (1)$$

in the celestial frame of reference. This includes the following two cases: (i) $H_{cosmic} = g\sigma \cdot \mathbf{n}$ with $C_1 = gn_1$, $C_2 = gn_2$, $C_3 = gn_3$ as considered in [14-16]; here C's are constants, and (ii) $H_{cosmic} = g\sigma \cdot \mathbf{v}$ with $C_1 = gv_1$, $C_2 = gv_2$, $C_3 = gv_3$ as considered in [12,13,17,18]; in this case, since $\mathbf{v}$ is largely, the velocity of our solar system through the cosmic preferred frame, to a first approximation, C's are also constants. For convenience, we use the celestial equatorial coordinate system with the earth rotation axis as z-axis and the direction of the spring equinox as the positive x-direction (Fig. 1). The right ascension $\alpha$ of our laboratory is measured eastward along the celestial equator from the spring equinox ($\Upsilon$) to its intersection with laboratory's hour circle. Declination $\delta$ is the geographical latitude. For our laboratory, $\delta$ is 24°47′43″ and the longitude is 120°59′58″. For a suspended electron polarized-body with its net spin axis pointing in a horizontal direction rotated counterclockwise from east direction by $\theta$, the torque from (1) is

$$\vec{\tau} = n\vec{C} \times <\vec{\sigma}>, \quad (2)$$

where $n$ is the number of polarized electrons and $<\vec{\sigma}>$ is the average polarization vector. When we suspend this polarized-body with a fibre, the torque on the fibre is

$$\begin{aligned}\tau_{vert} = &\frac{1}{2}n \mid <\vec{\sigma}> \mid [C_1(1+\sin\delta)\cos(\alpha+\theta) \\ &- C_1(1-\sin\delta)\cos(\alpha-\theta) + C_2(1+\sin\delta)\sin(\alpha+\theta) \\ &- C_2(1-\sin\delta)\sin(\alpha-\theta) - 2C_3\cos\delta\cos\theta].\end{aligned} \quad (3)$$

Table 1: Hughes-Drever types experiments using electron spins. $\delta E_\perp = 2(C_1^2 + C_2^2)^{1/2}$ and $\delta E_\parallel = 2 \mid C_3 \mid$ are the energy level splittings parallel and transverse to the earth rotation axis respectively. Of all the previous experiments, only [4] gives constraints on $\delta E_\parallel$.

| Reference | $\delta E_\perp$ ($10^{-18}$ eV) | $\delta E_\parallel$ ($10^{-18}$ eV) |
|---|---|---|
| Phillips(1987) [13] | $\leq 8.5$ | N.A. |
| Wineland et al. (1991) [4] | $\leq 550$ | $\leq 780$ |
| Chen et al. (1992) [14] | $\leq 7.3$ | N.A. |
| Wang et al. (1992)[15] | $\leq 3.87$ | N.A. |
| Chang et al. (1995) [16] | $\leq 2.96$ | N.A. |
| Berglund et al. (1995) [9] | $\leq 1.7$ | N.A. |
| This work | $\leq 0.057$ | $\leq 0.97$ |

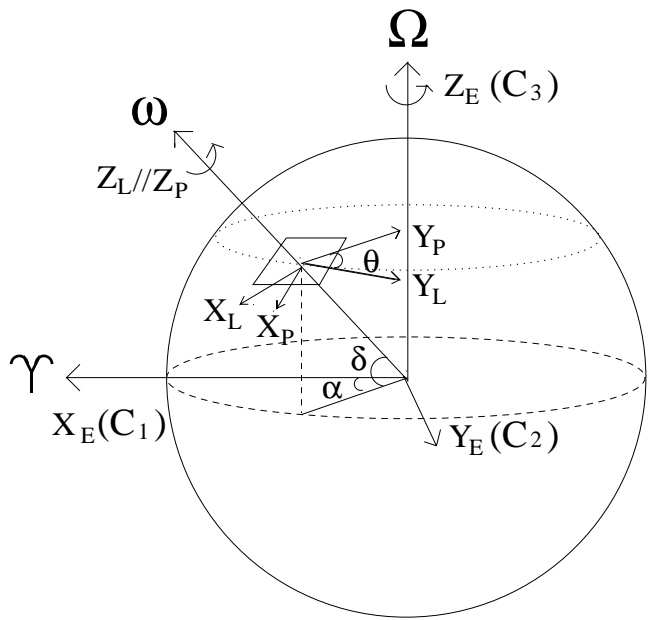

Figure 1: The celestial equatorial frame ($X_E$, $Y_E$, $Z_E$), the laboratory frame ($X_L$, $Y_L$, $Z_L$) and the rotatable table frame ($X_P$, $Y_P$, $Z_P$).

When the torsion pendulum is rotating with angular frequency $\omega$,

$$\theta = \omega t + \theta_0, \ \alpha = \Omega t + \alpha_0, \quad (4)$$

with $\Omega$ the earth-rotation angular frequency, $\theta_0$ the initial $\theta$ angle and $\alpha_0$ the initial right ascension.

The equilibrium angular position of the fibre is $\tau/K$ with $K$ the torque constant of the loaded fibre. In our experiment, we measure this angle position change to give constraints on $C_1$, $C_2$ and $C_3$. For $C_1$ and $C_2$, we use the earth rotation to perform 12-sidereal-hr offset substraction, and hence the accuracy in determining them is higher. We notice that the signals of $C_1$, $C_2$ and $C_3$ are at different frequencies — $\Omega + \omega$, $\Omega - \omega$ and $\omega$. For $C_3$, the constraint comes only from the frequency of the rotating table, and could not be observed from a non-rotating experiment. This is the new information from the rotating torsion-balance experiment, although its accuracy will not be so precise as those for $C_1$ and $C_2$. The measurement scheme will be presented in the measurement procedure following our description of various parts of the experimental setup (Fig. 2).

*The polarized body* — To obtain large net spins for increasing the possibility of detecting an anisotropy signal while still avoiding magnetic interaction, spin-polarized body of $Dy_6Fe_{23}$ was used in the previous experiments [19]. Dy-Fe compounds are ferrimagnetic at room temperature. The effective ordering of the iron lattice and dysprosium lattice have different temperature dependence because the strengths of exchange interactions are different. Near the compensation temperature, the magnetic moments of two lattices compensate each other mostly. $Dy^{+++}$ has $L = 5$ and $S = 5/2$. Half of the Dy magnetization comes from orbital angular momentum, the other half from spin. Most of the iron magnetization comes from spin. So there is a net spin (and net total angular momentum) remaining.

To make samples, $Dy_6Fe_{23}$ was synthesized by melting stoichiometric quantities of metallic iron and metallic dysprosium. The $Dy_6Fe_{23}$ ingots were crushed, pressed into a cylindrical aluminum cup, and magnetized along a tranverse direction. The magnetic moment of a small sample was measured as a function of temperature from 300 K to 4.2 K using an RF SQUID measurement system. We compare measurement data with model calculations to conclude that there is at least 0.4 net polarized electron per atom of $Dy_6Fe_{23}$. The net ferrimagnetic magnetization was shielded by two halves of pure iron casing, a thin aluminum spacer and a set of two fitting $\mu$-metal cups. The average magnetization after shielding is 2.57 mG ($4\pi$M).

Our $Dy_6Fe_{23}$ sample has diameter 16.0 mm, height 19.6 mm, mass 28.97 g, and number ($n \ |< \vec{\sigma} >|$) of net polarized electrons $8.95 \times 10^{22}$. The magnetically shielded polarized-body has a dimension of 22 mm $\phi \times$ 26 mm height with a mass 68.3 g.

*Torsion balance and rotatable table* — As in Fig. 2, the torsion balance is hung from the magnetic damper using a 25 $\mu$m General Electric tungsten fiber. The magnetic damper is hung from top of the chamber housing the torsion balance using a 75 $\mu$m tungsten fiber. The period of the torsion balance is measured to be 144.34 sec. The moment of inertia relative to the central vertical axis of the pendulum set is 32.32 g·cm$^2$. Hence, the torsion constant $K$ is $6.12 \times 10^{-2}$ dyne·cm/rad.

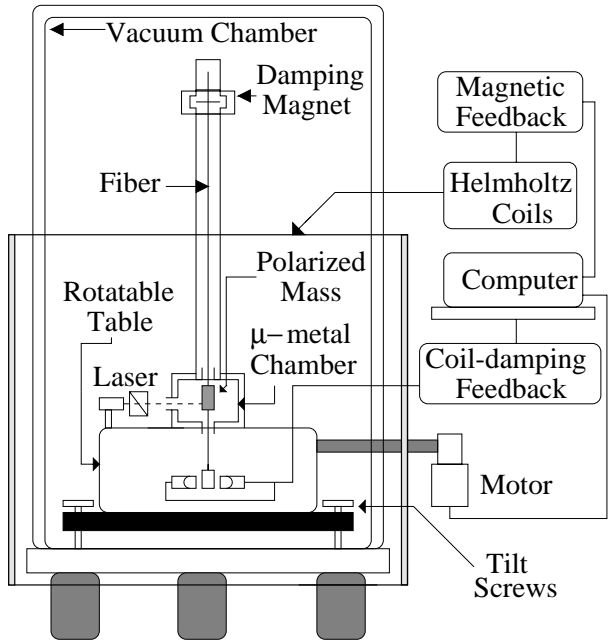

Figure 2: Schematic experimental set-up.

For the angle detection, we set up an optical level made of a 633 nm laser diode, a beamsplitter, a mirror on the torsion balance, a 30 cm focal-length cylindrical lens and a 3456-pixel Fairchild linear CCD with 7 $\mu$m pitch (1 pixel). 1 pixel difference in the CCD detection corresponds to 11.7 $\mu$rad deflection of the torsion balance and an amplitude of this size amounts to $4.98 \times 10^{-18}$ eV in the $C$'s. The torsional motion of the pendulum is damped by the D (Derivative) feedback of the coil damping system without changing the equilibrium position.

The torsion pendulum with its housing is mounted on a rotatable table fixed to a Huber Model 440 Goniometer. The angle positioning reproducibility is better than 2 arcsec and the absolute angle deviation is less than $\pm 10$ arcsec. The torsion balance together with the rotatable table and goniometer is mounted with 4 adjustable screws on the optical table inside the vacuum chamber. The four-phase stepping motor for rotating the table is outside the vacuum chamber and connected to the table by an acrylic rod. A 0.02 $\mu$rad resoultion biaxial tiltmeter is attached to the table to mointor the tilt.

*Magnetic field compensation and temperature control* — We use three pairs of square Helmholtz coils (1.2 m for each side) to compensate the earth magnetic field. A $\mu$-metal chamber with attenuation factor larger than 30 is placed inside the pendulum housing to magnetically shield the polarized-body. The magnetic field is measured 30 cm below the experiment chamber by a 3-axis magnetometer to make sure the region to be occupied by $\mu$-metal chamber and the polarized body to be less than 2 mG before we set up the torsion balance. The 3-axis magnetometer signals are fed back to control the currents of 3 pairs of Helmholtz coils with a precision better than 0.1 mG rms.

One thermometer is attached to the middle part of the aluminum tube housing the fiber. The other four are placed outside the wall of the vacuum chamber. The chamber temperature is controlled through an air conditioner and four radiant heater outside the chamber which are feedback controlled by these five thermometers through a personal computer. Temperature variation during 2-day data run is below 20 mK peak to peak for the thermometer on the tube.

*Measurement procedure* — Each complete data run consists of 4 contiguous periods. Each period lasts for 12 sidereal hours (11 hr 58 min 2 sec). In the first period we rotate the torsion balance clockwise or counterclcokwise with 1 hr period for 11 turns, and then stop the torsion balance for 58 min 2 sec to prepare for the second period. In the second period, we repeat with opposite rotation. In the third (fourth) period, we repeat with the same sense of rotation as in the second (first) period. The torsion balance angular position $F(t)$ is measured. $F(t)$ is basically equal to the equilibrium position $\tau/K$ plus deviation and noise. The signal part of $F(t)$, $\tau/K$, gives values of $C_1$, $C_2$, $C_3$, the deviation and noise gives uncertainty. Let $T$ be 12 sidereal hours. Adding two data sets with same rotating direction for $F$ using eqs. (3) and (4), we can eliminate $C_1$ and $C_2$ and estimate $C_3$; substracting, we can eliminate $C_3$ and estimate $C_1$ and $C_2$. For $0 \leq t \leq 12$ sidereal hours, in the case the rotation is counterclockwise in the first period, define $F_+(t) = F(t) - F(t + 3T)$ and $F_-(t) = F(t + 2T) - F(t + T)$; in the other case, define $F_+(t) = F(t + 2T) - F(t + T)$ and $F_-(t) = F(t) - F(t + 3T)$. We form the following combina-

tions to separate signals with different frequencies:

$$f_1(t) = \{(1+\sin\delta)F_+(t) + (1-\sin\delta)F_-(t)\}/(4\sin\delta)$$
$$= n|<\vec{\sigma}>|\{C_1\cos[(\Omega+|\omega|)t+\alpha_0+\theta_0]$$
$$+ C_2\sin[(\Omega+|\omega|)t+\alpha_0+\theta_0]\}, \quad (5)$$
$$f_2(t) = \{(1-\sin\delta)F_+(t) + (1+\sin\delta)F_-(t)\}/(4\sin\delta)$$
$$= n|<\vec{\sigma}>|\{C_1\cos[(\Omega-|\omega|)t+\alpha_0-\theta_0]$$
$$+ C_2\sin[(\Omega-|\omega|)t+\alpha_0-\theta_0]\}, \quad (6)$$
$$f_3(t) = \{F(t)+F(t+3T)\}/(2\cos\delta)$$
$$= -n|<\vec{\sigma}>|C_3\cos(|\omega|t+\theta_0), \quad (7)$$
$$f_4(t) = \{F(t+T)+F(t+2T)\}/(2\cos\delta)$$
$$= -n|<\vec{\sigma}>|C_3\cos(-|\omega|t+\theta_0). \quad (8)$$

*Analysis and results* — From the FFT analysis of the linear-drift-reduced CCD residuals of $f_1(t)$, $f_2(t)$, $f_3(t)$ and $f_4(t)$, we obtain two estimates of the $C_1$, $C_2$ and $C_3$. Fig. 3 shows a typical data set for $f_1(t)$ and its Fourier spectra. In this case, we strat rotating the torsion pendulum set at 22:58:07, February 16, 1999 counterclockwise for spin initially in the west direction ($\theta_0 = 180$). Because of starting transients, we discard the first hour data and use the interval $t = 1$ hr to $t = 10.599$ hr (10 cycles for angular frequency $\Omega + \omega$) for Fourier analysis. At $t = 1$ hr, the initial right ascension $\alpha_0$ is $279.6^o$. As we can see in Fig. 3(b), the 9th and 10th harmonics are higher than the neighboring harmonics. This is due to the contribution of uncancelled residue of one-hour rotation period (frequency = 0.2778 mHz) and is clear in Fig. 3(c) when we use ten-hour data for Fourier analysis. In Fig. 3(d), we substract this one-hour period residue from Fig. 3(b). The $\cos[(\Omega+\omega)t+\alpha_0+\theta_0]$ amplitude is now $-0.0018$ pixel and $\sin[(\Omega+\omega)t+\alpha_0+\theta_0]$ amplitude $-0.0028$ pixel, corresponding to $n|<\vec{\sigma}>|C_1 = 0.0031$ pixel and $n|<\vec{\sigma}>|C_2 = -0.0013$ pixel. An estimate of uncertainty is obtained by averaging the two neighboring FFT amplitudes with this amplitude; this gives an uncertainty of 0.0031 pixel. Converting to the estimate of $C_1$ and $C_2$, we have $(C_1^2+C_2^2)^{1/2} = (1.65 \pm 1.55) \times 10^{-20}$ eV. The accumulation of 16 days of data gives 16 sets of these number (8 sets for $f_1(t)$ and 8 sets for $f_2(t)$). The weighted average for $(C_1^2+C_2^2)^{1/2}$ is $(1.8\pm5.3)\times10^{-21}$ eV.

For the determination of $C_1$ and $C_2$, the effects with 1-hr rotation period are largely cancelled out in $F_+$ and $F_-$. The uncancelled residues in $f_1$ and $f_2$ can be substracted as explained in the last paragraph. However, for determination of $C_3$, the effects with rotation period need to be modelled in order to be able separate from the $C_3$ signals. The tilt effect is modelled. Other effects are put into systematic error. The later data have much less tilt effect. With these 8 days of data, $C_3$ is determined to be $(1.2\pm3.5)\times10^{-19}$ eV.

A better adjusted system for the tilt of the rotatable torsion balance is expected to reduce the noise. One order-of-magnitude improvement will reach the sensitivty to probe the macroscopic evidences of spin-rotation coupling $H_{eff} = -\vec{\Omega}\cdot\vec{S}$ on earth [20]. This non-inertial effect is calculated to be equivalent to a $C_3$ of $2.4 \times 10^{-20}$ eV.

Our constraint on the Lorentz and CPT violation parameters $\tilde{b}^e_\perp$ and $\tilde{b}^e_Z$ of Bluhm and Kostelecky is $\tilde{b}^e_\perp [= (C_1^2+C_2^2)^{1/2}] \leq 3\times10^{-29}$ GeV and $\tilde{b}^e_Z(=C_3) \leq 5\times10^{-28}$ GeV [21].

We thank Sheau-shi Pan, Wan-Sun Tse and Hsien-Chi Yeh for their help and encouragement. We also thank the National Science Council of the Republic of China for supporting this work.

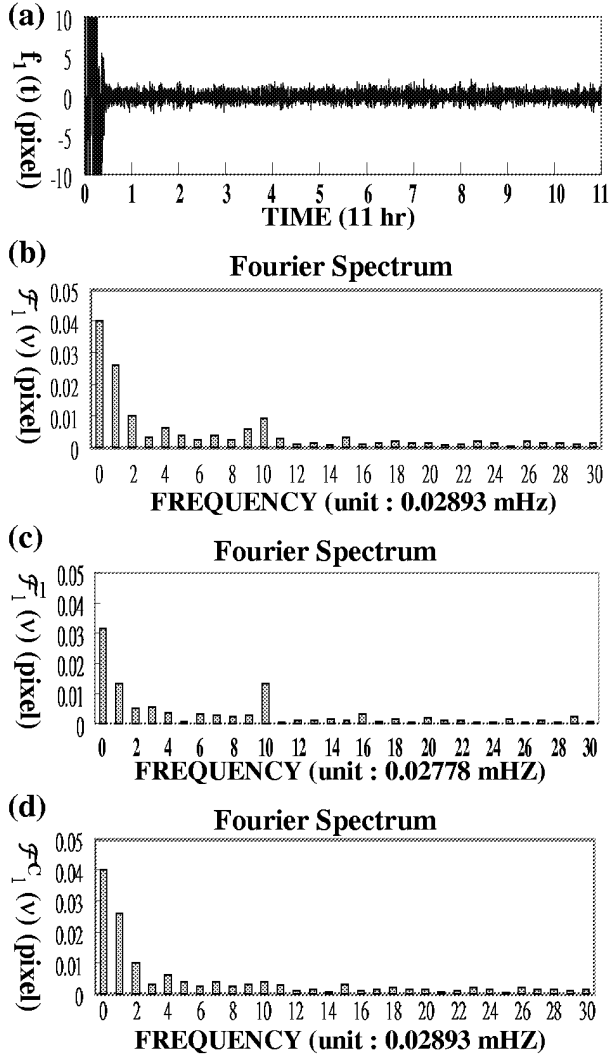

Figure 3: (a) A typical data set for $f_1(t)$. (b) Fourier spectrum $\mathcal{F}_1(\nu)$ for $f_1(t)$ from $t = 1$ hr to $t = 10.599$ hr. The first hour data is abandoned because of starting transients. The time interval for Fourier transform is exactly 10 cycles for angular frequency $\Omega + |\omega|$. (c) Fourier spectrum of 10 hr interval. The Fourier component with one-hour period is conspicuous. When this is substracted, the Fourier spectrum in (b) is corrected to (d).